\documentclass[prd,twocolumn,floats,superscriptaddress,eqsecnum,floatfix,showpacs]{
revtex4}

\usepackage{amssymb,amsmath}
\usepackage{epsfig}

\pdfoutput=1
\usepackage{amssymb,amsmath}
\usepackage{epsfig}
\usepackage{color}
\usepackage{datetime}

\newcommand\fverb{\setbox\fverbbox=\hbox\bgroup\verb}
\newcommand\fverbdo{\egroup\medskip\noindent%
			\fbox{\unhbox\fverbbox}\ }
\newcommand\fverbit{\egroup\item[\fbox{\unhbox\fverbbox}]}
\newbox\fverbbox

\newcommand{\be}{\begin{equation}}
\newcommand{\ee}{\end{equation}}
\newcommand{\ba}{\begin{eqnarray}}
\newcommand{\ea}{\end{eqnarray}}

\newcommand{\lap}{\bigtriangleup}

\newcommand{\eq}[1]{(\ref{#1})}

\newcommand{\bs}[1]{{\boldsymbol{#1}}}

\newcommand{\hh}{\, ,\hspace{1cm}}
\newcommand{\hhh}{\, ,\hspace{0.4cm}}

\newcommand{\ins}[1]{{\mbox{\tiny #1}}}

\newcommand{\inds}[1]{{\scriptscriptstyle #1}}

\begin{document}

\title{Bi-conformal symmetry and static Green functions in the
higher-dimensional Reissner-Nordstr\"{o}m spacetimes}

\author{Valeri P. Frolov\thanks{E-mail:
vfrolov@ualberta.ca}\, and
Andrei Zelnikov\thanks{E-mail: zelnikov@ualberta.ca}\\
Theoretical Physics Institute, Department of Physics\\
University of Alberta, Edmonton, AB, Canada T6G 2E1}

\begin{abstract}
We study a static scalar massless field created by a source located near an 
electrically charged
higher dimensional spherically symmetric black hole. We demonstrated that there exist 
bi-conformal
transformations relating static field solutions in the metric with different 
parameters of the mass
$M$ and charge $Q$. Using this symmetry we obtain the static scalar Green function in 
the higher
dimensional Reissner--Nordstr\"{o}m spacetimes.
\end{abstract}

\pacs{04.50.Gh, 04.40.Nr, 11.10.Kk}

\maketitle

\section{Introduction}

In this paper we continue studying  minimally coupled massless scalar fields created by static
sources placed in the vicinity of a higher dimensional static black holes. For this purpose we use
the method of bi-conformal transformations, which was developed and applied to the case of the
Schwarzschild-Tangherlini metrics in our previous paper \cite{Frolov:2014kia}. This 
method is based on the following observations.

A scalar
massless field $\varPhi$ in a  $D$-dimensional spacetime with metric $g_{\mu\nu}$
($\mu,\nu=0,\ldots, D-1$) obeys the equation
\be\label{box}
\Box \varPhi=-4\pi J\, .
\ee
Let us consider a static scalar field $\varPhi(X)$ created by a source $J(X)$ in the
static spacetime with the metric
\be\begin{split}\label{met}
&ds^2=-\alpha^2 dt^2+g_{ab}dx^a dx^b\,, \\
&X=(t,x^a)\hhh \alpha=\alpha(x)\hhh g_{ab}=g_{ab}(x)
\,.
\end{split}\ee
Then the equation (\ref{box}) is reduced and takes the form
\be\label{eqF}
\hat{F}\,\varPhi=-4\pi J\hh
\hat{F}={1\over\alpha\sqrt{g}}\partial_a \left(\alpha\sqrt{g}
g^{ab}\partial_b\right)\, .
\ee
Here $g=\det(g_{ab})$. The red-shift factor $\alpha$
 is connected with the norm of the static Killing vector $\bs{\xi}$ as follows: 
$\alpha=\sqrt{-\bs{\xi}^2}=\sqrt{-g_{tt}}$. The equation (\ref{eqF}) is
invariant under the following {\em bi-conformal} transformations
\be\label{bi-conf}
\varPhi=\bar{\varPhi}\, ,\ g_{ab}=\Omega^2 \bar{g}_{ab}\, ,\ \alpha=\Omega^{-n}\bar{\alpha}\, ,\
J= \Omega^{2}\bar{J}\, ,
\ee
where $n\equiv D-3$ and $\Omega$ is an arbitrary function of spatial coordinates 
$x^a$.

This transformation consists of a bi-conformal map 
\cite{GarciaParrado:2003hu,GomezLobo} of the original background 
$D$-dimensional metric 
$g_{\mu\nu}$
\be
\Psi_{\Omega}: \bs{g}\to \bar{\bs{g}}\, ,
\ee
accompanied by a properly chosen rescaling of the charge density $J$.
If one starts with a solution of the
Einstein equations, a new metric, obtained as a result of this transformation, is not 
necessarily
a solution of the  Einstein equations with a physically meaningful stress-energy 
tensor. However,
it may happen that for a specially chosen transformation this new metric has enhanced 
symmetries.

An interesting example is a Majumdar-Papapetrou metric, describing the gravitational 
field of a 
set 
of higher dimensional extremely charged black holes in equilibrium. Under properly 
chosen 
bi-conformal map this metric reduces to the higher dimensional Minkowski metric. This 
allows one to 
solve the static scalar field equation in the Majumdar-Papapertou exactly (see 
\cite{Frolov:2012jj}).

In the paper \cite{Frolov:2014kia} we demonstrated that the method of bi-conformal 
transformations can 
be used for solving static equations in spacetimes  of static spherically symmetric
black holes. The enhanced symmetry of the bi-conformal metric $\bar{\bs{g}}$ was used 
in that paper
to obtain the static Green functions for the equation (\ref{eqF}) in a higher
dimensional Schwarzschild-Tangherlini spacetime. In this paper we demonstrate how 
this method works 
for the case a charged higher dimensional  black hole.

There are many possible applications of the proposed result. One of them is an old 
problem of finding a 
self-energy and a self-force of charged particles
near black holes
\cite{Unruh:1976fc,Smith:1980tv,Zelnikov:1983,Zelnikov:1982in}. In four dimension the 
closed 
form of the exact  solution for the field of a point charges in the
black hole geometry was obtained earlier 
\cite{Zelnikov:1983,Zelnikov:1982in,Copson01031928,Linet:1976sq,Linet:1977vv,
Ottewill:2012aj}.

The recent interest to the problem of a self-force is stimulated by a study of
the back-reaction of the field on the particle moving near black holes 
\cite{Poisson:2011nh} in
connection with the gravitational wave emission by such particles.
More recently, several publications discussed higher-dimensional aspects of this
problem (see, e.g., \cite{Beach:2014aba,Frolov:2014gla}). This study was stimulated 
by general interest to 
spacetimes and brane
models with large extra dimensions.

The paper is organized as follows. In Section~\ref{section2} we discuss
bi-conformal transformations of higher dimensional spherically
symmetric metrics and demonstrate that the Reissner-Nordstr\"{o}m metrics
are bi-conformally related to the higher dimensional Bertotti-Robinson
metric. The latter is a product of 2D anti-de Sitter space and a
sphere. Using this result we construct a bi-conformal map of  the
Reissner-Nordstr\"{o}m metrics with different parameters of mass and
charge. In Section~\ref{section3} we obtain useful representations for  static
Green functions in a spacetime of static spherically symmetric higher
dimensional charged black holes. Section~\ref{section4} contains example of
calculations of the static Green functions for 4,5 and 6 dimensional
black holes. Section~\ref{section5} contains discussion of the obtained results and
their possible generalizations.

\section{Bi-conformal map and symmetry enhancement of static spherically symmetric
spacetimes}\label{section2}

\subsection{Symmetry enhancement condition}

Let us consider an application of the method of the bi-conformal maps to the case of  
a general 
static spherically symmetric $D$-dimensional metric. The corresponding metric is
\be\label{ds}
ds^2=-f(r)\,dt^2+w^{-1}(r)\,dr^2+r^2\,d\omega^2_{n+1}\, ,
\ee
where $n=D-3$ and  $d\omega_{n+1}^2$ is the line
element on a $(n+1)$-dimensional unit sphere
\be\label{theta}
d\omega_{n+1}^2=d\theta_{n}^2+\sin^2\theta_{n}\,d\omega_{n}^2\hh
d\omega_{0}^2 = d\phi^2\, .
\ee
We denote $\theta_0\equiv\phi\in[0,2\pi]$. All other coordinates
$\theta_{i>0}\in[0,\pi]$.
This metric is invariant under  time translations and spatial rotations.
Since $ds$, $t$ and $r$ have the same dimensionality of the length, the metric 
(\ref{ds}) can be 
presented in the  form $ds^2=a^2 dS^2$, where the dimensionless metric $dS^2$ is 
obtained from  
(\ref{ds}) by substituting $t\to t/a$ and $r\to r/a$,  where $a$ is an arbitrary 
constant parameter 
with the dimensionality of length.

Let us apply a bi-conformal transformation (\ref{bi-conf}) to this metric with 
$\Omega=r/a$. This 
choice guarantees that $\Omega$ is dimensionless. After this bi-conformal 
transformation one has
\be\begin{split}\label{dsdh}
d\bar{s}^2&=dh^2+a^2 \,d\omega^2_{n+1}\,,\\
dh^2&=-\left({r\over a}\right)^{2n}f(r)\,dt^2+{a^2\over r^2
w(r)}\,dr^2\, .
\end{split}\ee
The scalar curvature of the two-dimensional metric $dh^2$ is
\be\begin{split}\label{R}
R&= -{1\over 2 a^2 f^2}\left\{rfw'(2nf+rf')\right.\\
&\left.+w\,[
2r^2f''-r^2(f')^2+2r(2n+1)ff'+4n^2f^2]\right\}\, .
\end{split}\ee
Here and later $(\ldots)'=d(\ldots)/dr$. The metric $dh^2$ possesses an enhanced 
symmetry if its 2D
curvature $R$ is constant. We denote its value by
\be\label{RR}
R=-{2\over b^2}\, ,
\ee
where $b$ is a constant of the dimensionality of the length. The equations (\ref{R}) 
and (\ref{RR})
can be solved to determine the function $w$. The result is
\be\label{w}
w=\left( {a^2\over n^2 b^2}+{C\over r^{2n} f}\right) \left(1+{rf'\over 2 n 
f}\right)^{-2}\, .
\ee
Here $C$ is an integration constant. Let us suppose that function $f$ has the 
following asymptotic
at infinity
\be
f=f_0+f_1 r^{-\gamma}+\ldots\hh \gamma\ge 1\, .
\ee
Then \eq{w} shows that asymptotic value of $w$ at the infinity is $a^2/(n^2 b^2)$. 
The spacetime
does not have a solid angle deficit and is asymptotically flat only if
\be\label{abn}
{a\over n b}=1\, .
\ee
In what follows we always assume that this conditions is satisfied.

By using the relation (\ref{w}) one finds such functions $\{f(r),w(r)\}$, for which 
the bi-conformal
transformation of the metric (\ref{ds}) has an enhanced symmetry. The corresponding 
metric
\be
d\bar{s}^2=dh^2+a^2 d\omega^2_{n+1}
\ee
is a direct sum of the two dimensional anti-de Sitter metric $dh^2$ and the metric
$a^2 d\omega^2_{n+1}$ on $(n+1)$-dimensional sphere. The ratio of the curvature radii 
for these two
metrics is fixed by the condition \eq{abn}. This metric describes a particular 
Bertotti-Robinson 
spacetime and can be written in the following canonical form
\be\label{B-R}
d\bar{s}^2=a^2\left[{1\over n^2}\left(-(\rho^2-1)\,d\bar{\sigma}^2+{1\over
\rho^2-1}\,d\rho^2\right) +d\omega^2_{n+1}\right] \,.
\ee
Let us emphasize that the parameter $a$ has dimensionality of the length and it is 
arbitrary.

\subsection{Bi-conformal map of Reissner-Nordstr\"{o}m metric to the 
Bertotti-Robinson space}

Let us consider a special case of the metric (\ref{ds}) with an extra condition
\be\label{wfe}
w=f\, .
\ee
For this choice the relation (\ref{w}) becomes an equation which allows one to obtain 
the function 
$f$.
The ordinary differential equation (\ref{w}) with $w=f$ is of the first order. Hence 
its solution 
besides the constant $C$ contains
another arbitrary integration constants $C_1$. It is possible to show that one can 
choose these 
constants so that the solution takes the form
\be\label{RN}
f=1-{2M\over r^n}+{Q^2\over r^{2n}}\, .
\ee
For real positive $M$ and real $Q$, which satisfies the condition $|Q|\le M$, the 
metric (\ref{ds}) 
with (\ref{wfe}) and (\ref{RN}) is the metric of a higher dimensional spherically 
symmetric 
electrically charged black hole with $M$ and $Q$ being its mass and charge, 
respectively.

In order to rewrite the metric $d\bar{s^2}$, obtained as a result of the bi-conformal 
map 
(\ref{dsdh}), in the standard (canonical) form (\ref{B-R}) it is sufficient to make 
the following 
coordinate transformations
\be\label{rho_r}
r^n=M+\mu\rho\hhh t={a^{n+1}\over n\mu}\,\bar{\sigma}  \hhh
\mu=\sqrt{M^2-Q^2}\, .
\ee
We denote this bi-conformal map as follows
\be\label{bc}
\Psi_{\Omega}: \bs{g}_{M,Q}\to \bar{\bs{g}}_{BR}\hhh
\Omega=r/a \, .
\ee

\subsection{Bi-conformal transformations within the Reissner-Nordstr\"{o}m family of 
solutions}

The method of bi-conformal maps was used in the paper \cite{Frolov:2014kia} to 
obtain static Green 
functions 
in the background of the higher dimensional Schwarzschild-Tangherlini spacetimes. 
For 
this purpose, 
one uses at first the enhanced symmetry of a related Bertotti-Robinson space to find 
the 
$D$-dimensional Green function in this space, and after this one obtains the static 
Green function 
by means of the dimensional reduction. One can apply the same method for finding 
static Green 
functions in the  Reissner-Nordstr\"{o}m geometry. However, there exist another much 
simper way. One 
can generate the corresponding static Green function in the spacetime of charged 
black holes by 
using the already known Green function for the Schwarzschild-Tangherlini spacetime.

For this purpose let us notice that the canonical form (\ref{B-R}) is  {\em 
universal} in the 
following sense: It is the same for any Reissner-Nordstr\"{o}m
metric and it does not depend on its parameters $M$ and $Q$. This observation opens 
an interesting
possibility to relate metrics with different parameters. Let us introduce new 
coordinates
$\hat{t}$ and $\hat{r}$
\be
\hat{r}^n=\hat{M}+\hat{\mu}\rho  \hhh \hat{t}={a^{n+1}\over 
n\hat{\mu}}\,\bar{\sigma}\hhh 
\hat{\mu}=\sqrt{\hat{M}^2-\hat{Q}^2}\, ,
\ee
and denote
\be
\hat{\Omega}=\hat{r}/a \, .
\ee
Then one has the following bi-conformal map of the Reissner-Nordstr\"{o}m
metric with parameters $\hat{M}$ and $\hat{Q}$ to the canonical Bertotti-Robinson 
metric
\be\label{bch}
\Psi_{\hat{\Omega}}: \bs{g}_{\hat{M},\hat{Q}}\to \bar{\bs{g}}_{BR}\, .
\ee
Combining the direct bi-conformal map (\ref{bc}) with the bi-conformal map, inverse 
to (\ref{bch}), 
one obtain a bi-conformal map
\be\label{ppsi}
\Psi=\Psi^{-1}_{\hat{\Omega}}\circ \Psi_{\Omega}: \bs{g}_{M,Q}\to 
\bs{g}_{\hat{M},\hat{Q}}\, .
\ee
This bi-conformal map is a transformation of the original Reissner-Nordstr\"{o}m 
metric with 
parameters ${M}$
and ${Q}$ to a similar metric with different parameters  $\hat{M}$
and $\hat{Q}$. The static equation (\ref{eqF}) is invariant under such a 
transformation provided
one in addition properly transforms  the source term $J\to \hat{J}$.

In other words the solutions for the static field $\varPhi$ in the original 
background
space are simply related to solutions in a spacetime with modified parameters of the 
mass and
the charge. In particular, if one knows the static Green function in the spacetime of 
uncharged 
black hole, one can obtain the static Green function for
the charged black hole by using the above described transformations. In the next 
section we 
demonstrate how this method works in more detail.

\section{Static Green functions}\label{section3}

\subsection{Bi-conformal map of static Green functions}

Following the paper \cite{Frolov:2014kia} we define a static Green function 
${G}(x,x')$ as follows
\be\label{Gret}
{G}(x,x')=\int_{-\infty}^{\infty} dt\,\mathbb{G}_\ins{Ret}(t,x;0,x')\, .
\ee
Here $\mathbb{G}_\ins{Ret}(t,x;0,x')$ is a retarded Green function in the 
D-dimensional spacetime. 
This static Green function satisfies the the equation
\be\label{FG}
\hat{F}\,{G}(x,x') = -{\delta(x-x')\over\alpha\sqrt{g}}\, .
\ee
In what follows, we assume that this Green function is decreasing when one of its 
parameters $x$ 
tends to infinity and remains regular at the horizon (for more details see
\cite{Frolov:2014kia}).

The static Green function is simply related to the expression for a scalar field 
created by a point 
charge. The current of a static point charge $q$ positioned at the point $y$ reads
\be
J(x)=q{\delta(x-y)\over \sqrt{g}}\,.
\ee
In this case the scalar field at the point $x$ takes the form\footnote{Note that 
rescaling of the time variable by a constant factor $t=c\tilde{t}$ leads to 
$\tilde{\alpha}=c\alpha$,  $\tilde{J}=J $, $\tilde{G}(x,x')=c^{-1}G(x,x')$,
$\tilde{\varPhi}=\varPhi\,.$
Thus the scalar field $\varPhi$ created by the static source is invariant under this 
constant
rescaling of the time coordinate.
}
\be\label{potential}
\varPhi(x)=4\pi q\,\alpha(y)\,{G}(x,y)\,.
\ee
The field of a distributed source $q(y)$ can be easily obtained by integration over 
$y$ of the 
right-hand side of this relation.

It is convenient to introduce a new radial variable $\rho$ related to the
radial coordinate $r$ as follows \eq{rho_r}
\be
\rho={r^n -M\over \mu}\, .
\ee
The Reissner-Nordstr\"{o}m metric takes the form
\be\begin{split}\label{R-N}
ds^2&=-{\mu^2(\rho^2-1)\over(M+\mu\rho)^2}\,dt^2\\
&+(M+\mu\rho)^{2/n}\left[{1\over
n^2(\rho^2-1)}\,d\rho^2+d\omega^2_{n+1} \right]
\,.
\end{split}\ee
The horizon corresponds to
$
\rho=1
$
and the gravitational radius $r_\ins{g}$ is given by the expression
$
r_\ins{g}^n=M+\mu\,.
$
The surface gravity at the horizon is
\be\label{kappa}
\kappa={n\mu \over r_\ins{g}^{n+1}}\,.
\ee
In these coordinates the equation for the static Green function takes the form
\be\begin{split}\label{Reissner_Lap}
\left[n^2(\rho^2-1)\,\partial^2_\rho+2n^2\rho\,\partial_\rho+\lap_{\omega }^{n+1}
\right] G(x,x')&\\
=-{n\over \mu}\,\delta(\rho-\rho')\delta(\omega ,\omega')\,.&
\end{split}\ee
Here $\lap^{n+1}_{\omega}$ and $\delta(\omega ,\omega')$ are the Laplace
operator and a covariant delta-function on the unit $(n+1)$-dimensional sphere,
respectively,
\be\begin{split}
&\lap^{n+1}_{\omega}=\partial^2_{\theta_{n}}+n{\cos\theta_{n}
\over\sin\theta_{n}}
\partial_{\theta_{n}}+{1\over\sin^2\theta_{n}}\lap^{n}_{\omega}\,,\\
&\lap^{1}_{\omega}=\partial^2_{\phi}\, ,\\
&\delta^{n+1}(\omega ,\omega'={\delta(\theta_{n}-\theta'_{n})\over
\sin^n\theta_{n}}\delta^{n}(\omega ,\omega')\,,\\
&\delta^{1}(\omega ,\omega')=\delta(\phi-\phi')\, .
\end{split}\ee

Because of the spherical symmetry of background geometry, the resulting static Green 
functions
are the functions of radial coordinates of the observer $\rho$, the source $\rho'$, 
as well as the angular
distance $\gamma\equiv\gamma_{\inds{n+1}}$ between the source and the observational 
point.
\be
G(x,x')=G(\rho,\rho';\gamma)\,.
\ee
The  angular distance on the $(n+1)$ dimensional sphere can be written explicitly in 
terms of the
angular coordinates \eq{theta}
\be\begin{split}
&\cos\gamma_{n+1}=\cos\theta_{n}\cos\theta'_{n}+\sin\theta_{n}
\sin\theta'_{n}\cos\gamma_{n}\,,\\
&\gamma_{\inds{0}}=\phi-\phi'\,.
\end{split}\ee

The canonical Bertotti-Robinson spacetime \eq{B-R} is homogeneous.
In the paper \cite{Frolov:2014kia} we have used the knowledge of heat kernels on 
homogeneous 
spaces to
derive the static Green functions. Now, using the bi-conformal symmetry of the static 
operator
\eq{eqF}, we can use these results to derive the static Green functions in the
Reissner-Nordstr\"{o}m spacetime with arbitrary parameters of the mass $M$ and the 
charge $Q$.

One can see that the bi-conformal transformation \eq{bi-conf} with
\be\label{Omega_a}
\Omega={r\over a}={\left(M+\mu\rho\right)^{1/n}\over a}\,,
\ee
leads to the Bertotti-Robinson canonical metric \eq{B-R},
if $\bar{\sigma}$ is identified with the rescaled Reissner-Nordstr\"{o}m time 
coordinate $t$
\be\label{bar_sigma}
\bar{\sigma}=\bar{\kappa}t \hhh 
\bar{\kappa}={n\mu\over 
a^{n+1}}=\left({r\ins{g}\over 
a}\right)^{n+1}\kappa\,.
\ee
Here $\kappa$ is given by \eq{kappa} and $\bar{\kappa}$ is the surface gravity of 
the 
horizon 
$\rho=1$ in the Bertotti-Robinson
spacetime, normalized according to the Killing vector 
$\bar{\xi}^{\mu}=\delta^{\mu}_t$
\be
\bar{\kappa}^2=-{1\over 
2}\bar{\xi}^{\alpha;\beta}\bar{\xi}_{\alpha;\beta}\Big|_\inds{\rho=1}\,.
\ee
One can define the static Green function in the canonical metric \eq{B-R} as the 
integral over the 
dimensionless time coordinate $\bar{\sigma}$
\be\label{GB-R}
\bar{G}(x,x')=\int_{-\infty}^{\infty} 
d\bar{\sigma}\,\bar{\mathbb{G}}_\ins{Ret}(\bar{\sigma},x;0,x')\,,
\ee
where $\bar{\mathbb{G}}_\ins{Ret}(\bar{\sigma},x;\bar{\sigma}',x')$ is a retarded 
Green function in 
the canonical Bertotti-Robinson spacetime \eq{B-R}. 
It satisfies the the equation
\be\begin{split}\label{BR_Lap}
\left[n^2(\rho^2-1)\,\partial^2_\rho+2n^2\rho\,\partial_\rho+\lap_{\omega }^{n+1}
\right] \bar{G}(x,x')&\\
=-{n^2\over a^{n+1}}\,\delta(\rho-\rho')\delta(\omega 
,\omega')\,.&
\end{split}\ee
The left-hand side of this equation coincides with that of
\eq{Reissner_Lap}. The right-hand sides of these equations differ only by a constant 
factor related 
to the rescaling of the time coordinate \eq{bar_sigma}. Thus, the static Green
functions in these spaces also differ only by a constant factor
\be\label{GbarG}
G(\rho,\rho';\gamma)={1\over \bar{\kappa}}\bar{G}(\rho,\rho';\gamma)\,.
\ee

To construct the bi-conformal map (\ref{ppsi}) relating Reissner-Nordstr\"{o}m with 
different 
parameters $M$ and $Q$ one proceeds as follows.
Let us define two radial coordinates $r$ and $\hat{r}$ by the relation
\be\label{rr}
{r^n -M\over\mu}={\hat{r}^n -\hat{M}\over \hat{\mu}}\equiv \rho\,.
\ee
Here
\be
\mu=\sqrt{M^2-Q^2}\hhh \hat{\mu}=\sqrt{\hat{M}^2-\hat{Q}^2}\, .
\ee
This allows one to express the new coordinate $\hat{r}$ in terms of the original 
radial coordinate 
$r$. The time coordinates are related as follows
\be
\hat{t}={\mu\over \hat{\mu}}\,t\, .
\ee

The the bi-conformal transformation with
\be\label{Omega_11}
\Omega=\left[{M+\mu\rho\over
\hat{M}+\hat{\mu}\rho}\right]^{1/n}
\ee
relates two arbitrary Reissner-Nordstr\"{o}m metrics \eq{R-N} characterized by the 
parameters
$M,Q$ and $\hat{M},\hat{Q}$, correspondingly.

Because of the time rescaling the relation between the
static Green functions in these Reissner-Nordstr\"{o}m spacetimes becomes
\be\label{GG}
\mu\,G(r,r';\gamma)=\hat{\mu}\,\hat{G}(\hat{r},\hat{r}';\gamma)\,.
\ee

Note that though the static Green function depends on the time rescaling, this 
dependence is dropped 
out of the expression for the scalar field $\varPhi$. The resulting $\varPhi$ is 
invariant with 
respect to
the time rescaling. One can say that the static scalar potentials $\varPhi$ for
all Reissner-Nordstr\"{o}m geometries are given by the same function of $\rho$. In
terms of the radial coordinates $r$ and $\hat{r}$ they are related by the coordinate
transformation \eq{rr}. Therefore, as soon as we know the static scalar Green
function for a particular choice
of the charge of a black hole, for example, for a neutral one, the identity
\eq{GG} makes it possible to generate the solution for the scalar field near
the Reissner-Nordstr\"{o}m black hole with an arbitrary mass and charge.

Since the cases of even and odd-dimensional spacetimes differ, so that we shall treat 
them 
separately.

\subsection{Even-dimensions}

In even dimensions the exact static Green function can be represented in the form of 
the
integral

\be\label{EvenG}
{G}(x,x')={1\over n\mu}\,{1\over 2\,(2\pi)^{n+3\over 2}}
\left({\partial\over \partial \cos\gamma}\right)^{(n+1)/2}
\,\int_0^{2\pi}d\sigma\,A_n \,.
\ee
Here $n=D-3$ and
\be\begin{split}\label{chin}
\cosh(\chi)&=\rho\rho'-\sqrt{\rho^2-1}\sqrt{\rho'{}^2-1}\cos\sigma\,.
\end{split}\ee
When $n\ge 2$, the functions $A_n(\sigma,\rho,\rho';\gamma)$ are given by the 
integral
\be\begin{split}\label{A_n}
A_n&=\int_{\chi}^{\infty}dy\,{1\over\sqrt{\cosh\left({y}\right)-\cosh\left({
\chi}\right)}}\,
{\sinh\left({y\over n}\right)\over
\sqrt{\cosh\left({y\over n}\right)-
\cos\left({\gamma}\right)}} \,.
\end{split}\ee
At large $y$ the integrand in \eq{A_n} behaves like $\exp[-y(n-1)/(2n)]$. Therefore, 
\eq{A_n} is convergent for any $n\ge 2$. In the case of the four-dimensional 
spacetime $(n=1)$ the integrand has to be modified to guarantee convergence of the 
integral. For example, one can subtract the asymptotic of the integrand, which does 
not depend on $\gamma$. Since \eq{EvenG} contains the derivative 
of $A_n$ over $\gamma$, the resulting Green function does not depend on the 
particular form of the subtracted $\gamma$-independent asymptotic. Thus, 
for $n=1$ one can choose
\be\begin{split}\label{A_1}
A_1&=\int_{\chi}^{\infty}dy\,{\sinh\left({y}\right)\over\sqrt{\cosh\left({y}
\right)-\cosh\left({\chi}\right)}}\\
&\times\left[{1\over\sqrt{\cosh\left({y}\right)-\cos\left({\gamma}\right)}} 
-{1\over\sqrt{\cosh\left({y}\right)+1}}\right]\,.
\end{split}\ee
Substitution 
\be\label{rhor}
\rho={r^n -M\over\mu}={r^n-M\over\sqrt{M^2-Q^2}}
\ee
into \eq{EvenG} gives the static Green function of a scalar
charge near the Reissner-Nordstr\"{o}m black hole \eq{R-N} in terms of the radial 
coordinate $r$.

\subsection{Odd-dimensions}

In odd-dimensional spacetimes we have
\be\label{OddG}
{G}(x,x')={1\over n\mu}\,{1\over \sqrt{2}\,(2\pi)^{{n+4\over 2}}}
\left({\partial\over \partial
\cos\gamma}\right)^{n/2}\,\int_0^{2\pi} d\sigma\,B_n\,,
\ee
where $n=D-3$ and $\chi$ is given by \eq{chin} and
\be
B_n=\int_{\chi}^{\infty}dy\,{1\over\sqrt{\cosh 
y-\cosh\chi}}\,{\sinh\left({y\over n}\right)\over
\cosh\left({y\over n}\right)-\cos\gamma} \,.
\ee

Similarly to the even dimensions, the Green function in the radial coordinates $r$
can be obtained after the substitution \eq{rhor}.

\section{Closed form of the Green function: Examples}\label{section4}

\subsection{Four dimensions. $D=4$}

In four dimensions ($n=1$) the integral \eq{A_1} can be done and one obtains
\be
A_1=\ln\left({\cosh\left(\chi\right)+1\over
\cosh\left(\chi\right)-\cos\left(\gamma\right)} \right) \,.
\ee
The integral over $\sigma$ can be taken explicitly and we obtain the closed form for 
the
static Green function
\be\begin{split}\label{G4}
G(x,x')&={1\over 4\pi \mu}
{1\over \sqrt{ \rho^2+\rho'{}^2-2\rho\rho'\cos\gamma
-1+\cos^2\gamma }} \,, \\
\rho&={r-M\over\mu}={r-M\over \sqrt{M^2-Q^2}}\,.
\end{split}\ee
When written in terms of the radial coordinate $r$ it reads
\be\begin{split}
G(x,x')={1\over 4\pi \mathcal{R}}\,,
\end{split}\ee
where
\be\begin{split}
\mathcal{R}^2&=(r-M)^2+(r'-M)^2\\
&-2(r-M)(r'-M)\cos\gamma
-(M^2-Q^2)\sin^2\gamma \,.
\end{split}\ee
This formula exactly reproduces the closed form of the well known result for
the scalar Green function in four-dimensional Schwarzschild geometry
\cite{Linet:1977vv,FrolovZelnikov:1980,Zelnikov:1982in}.
It is easy to check that using the bi-conformal symmetry \eq{GG} this solution could 
be generated
from that of the Schwarzschild case ($Q=0$).

In the limit of the extremally charged black hole $Q=M$ the obtained solution \eq{G4} 
reproduces
the result \cite{Frolov:2012jj} for the four-dimensional Majumdar-Papapetrou 
geometry.

\subsection{Five dimensions. $D=5$}

The other case, when there exists a closed  form for the static Green function is 
five-dimensional
($n=2$) Reissner-Nordstr\"{o}m black hole. One can generate this solution using the 
bi-conformal
symmetry \eq{GG} from that of the Tangherlini black hole \cite{Frolov:2014kia}, or, 
equivalently, 
just make
the substitution \eq{rhor} in the expression for the five-dimensional Green function 
(see eq.(6.14)
of \cite{Frolov:2014kia}).

\be\begin{split}\label{5DGreen}
G(x,x')&={1\over 8\pi^2 \mu}\,{1\over(\rho^2-1)^{1/4}(\rho'{}^2-1)^{1/4}}\\
&\times{\partial\over \partial \cos\gamma}\left\{
\,\varkappa\,\left[\mathbf{F}\left(\psi,\varkappa\right)+\mathbf{K}
\left(\varkappa\right)\right]\right\}\,,
\end{split}\ee
where $\mathbf{F}$ and $\mathbf{K}$ are the elliptic functions 
\be
\rho={r^2-M\over\mu}\,,
\ee
and
\be\begin{split}
&\sin\psi=\cos\gamma\,{\sqrt{2}\over
\sqrt{\rho\rho'-\sqrt{\rho^2-1}\sqrt{\rho'{}^2-1}+1}}  \,,
\\
&\varkappa={\sqrt{2}\,(\rho^2-1)^{1/4}(\rho'{}^2-1)^{1/4}
\over\sqrt{\rho\rho'+\sqrt{\rho^2-1}\sqrt{\rho'{}^2-1}+1-2\cos^2\gamma}} \,.
\end{split}\ee

To the best of our knowledge, this closed form for the static Green function in 
five-dimensional
Reissner-Nordstr\"{o}m black hole is new.

In the limit of the extremally charged black hole, when $Q=M$, the expression 
\eq{5DGreen} leads to
\be
\label{5DGreenExtreme}
G(x,x')=\displaystyle{1\over 4\pi^2 \mathcal{R}^2}\,,
\ee
where
\be\begin{split}\nonumber
\mathcal{R}^2&=(r^2-M)+(r'{}^2-M)\\
&-2\cos\gamma\sqrt{r^2-M}\sqrt{r'{}^2-M}.
\end{split}\ee
It exactly reproduces the result \cite{Frolov:2012jj} for the five-dimensional 
Majumdar-Papapetrou
geometry in the case of a single extremal black hole of the mass $M$.

\subsection{Six dimensions. $D=6$}

Application of the \eq{EvenG} to six-dimensional ($n=3$) Reissner-Nordstr\"{o}m black 
hole leads to
\be\begin{split}
A_3&=\int_{\chi}^{\infty}dy\,{1\over(\cosh
y-\cosh\chi)^{1/2}}\,{\sinh\left({y\over 3}\right)
\over\sqrt{\cosh\left({y\over 3}\right)-
\cos\left({\gamma}\right)}}\\
&=3\int_{\cosh(\chi/3)}^{\infty}dz\,{1\over\sqrt{4z^3-3z-\cosh\chi}}
\,{1\over\sqrt{z-\cos\gamma}} \,.
\end{split}\ee
This integral can be expressed in terms of the elliptic function $\mathbf{F}$
\be
A_3={6\over 
\sqrt{v(w-u)}}\mathbf{F}\left(\arcsin\sqrt{w-u\over 
w},{w(v-u)\over v(w-u)}\right)\,,
\ee
where
\be\begin{split}
&p=\cosh(\chi/3)\,, \hskip 1.21cm
w=2(p-\cos\gamma)\,,\\
&u=3p-i\sqrt{3p^2-3}\hhh v=3p+i\sqrt{3p^2-3}\,.
\end{split}\ee
Note that $A_3$ is real in spite of the complexity of the functions $u$ and $v$.
Thus the static Green function in the six-dimensional Schwarzschild-Tangherlini 
spacetime is given by the integral
\be
G(x,x')={1\over 48  \pi^3 \mu}
\left({\partial\over \partial \cos\gamma}\right)^{2}
\,\int_0^{2\pi}d\sigma\, A_3 \,.
\ee

It is problematic to obtain an answer for the Green functions in a closed form
for $D\ge 6$. However, a rather simple integral representation is possible in all
higher dimensions. For some applications, like computing of the
self-force and self-energy of scalar charges this integral representations is 
sufficient to obtain 
the final results in a closed form.

%%%%%%%%%%%%%%%%%%%%%%%%%%%%%%%%%%%%%%

\section{Discussion}\label{section5}

In this paper we demonstrated that there exist bi-conformal transformations relating 
static
solutions of the minimally coupled massless field equation in the 
Reissner-Nordstr\"{o}m spacetimes
with different values of the parameters of the mass $M$ and the charge $Q$. We used  
this symmetry
to generate expressions for the static Green functions in such space starting from 
similar Green
functions for the neutral (uncharged) higher dimensional black holes, that have been 
obtained
earlier \cite{Frolov:2014kia}. To check the obtained results, we considered limit of 
higher 
dimensional extreme
black holes with $|Q|=M$. This is a special case of the Majumdar-Papapetrou metrics 
related by
means of a bi-conformal map to the flat spacetime. It is possible to show that the 
obtained static Green
functions in a generic Reissner-Nordstr\"{o}m spacetime obey a correct flat spacetime 
limit.

Natural applications of the results obtained in our earlier publication 
\cite{Frolov:2014kia} and in 
this paper
is study of the problem of the self-energy and self force of point scalar charged in 
the background
of higher dimensional static black holes. Especially interesting is the origin of 
near horizon
logarithmic terms in these expressions in odd dimensional black holes 
\cite{Beach:2014aba,Frolov:2014gla} and the
relation of these terms with the bi-conformal anomalies (see discussion in
\cite{Frolov:2012xf,Frolov:2012ip,Frolov:2013qia}). It is interesting also to test 
the method of the
bi-conformal transformations in application to electric fields of static sources in 
the static black
hole backgrounds. Another interesting question is: Is it possible to generalize the 
method of
bi-conformal to the case of  fields from stationary sources in a spacetime of 
rotating black holes.
We are going to address these questions in our further work.

%%%%%%%%%%%%%%%%%%%%%%%%%%

\acknowledgments{
This work was partly supported  by  the Natural Sciences and Engineering
Research Council of Canada. The authors are also grateful to the Killam Trust
for its financial support.}

%%%%%%%%%%%%%%%%%%%%%%%%%%%%%%%%%%

%\bibliography{references}{}
%\bibliographystyle{h-physrev4}

%\begin{thebibliography}{99}

%\end{thebibliography}

\end{document}